# Distributed Architectures and Constellations for γ-Ray Burst Science

Fabrizio Fiore [1,*], Norbert Werner [2] and Ehud Behar [3]

1. INAF-Osservatorio Astronomico di Trieste, via Tiepolo 11, 34143 Trieste, Italy and Institute for the Fundamental Physics of the Universe, IFPU, via Beirut, 2, 34151 Trieste, Italy
2. Department of Theoretical Physics and Astrophysics, Faculty of Science, Masaryk University, Kotlářská 2, Brno 611 37, Czech Republic; werner@physics.muni.cz
3. Department of Physics, Technion, Haifa 32000, Israel; behar@physics.technion.ac.il
* Correspondence: fabrizio.fiore@inaf.it

**Abstract:** The gravitational wave/γ-ray burst GW/GRB170817 event marked the beginning of the era of multi-messenger astrophysics, in which new observations of Gravitational Waves (GW) are combined with traditional electromagnetic observations from the very same astrophysical source. In the next few years, Advanced LIGO/VIRGO and KAGRA in Japan and LIGO-India will reach their nominal/ultimate sensitivity. In the electromagnetic domain, the Vera C. Rubin Observatory and the Cherenkov Telescope Array (CTA) will come online in the next few years, and they will revolutionize the investigation of transient and variable cosmic sources in the optical and TeV bands. The operation of an efficient X-ray/γ-ray all-sky monitor with good localisation capabilities will play a pivotal role in providing the high-energy counterparts of the GW interferometers and Rubin Observatory, bringing multi-messenger astrophysics to maturity. To reach the required precision in localisation and timeliness for an unpredictable physical event in time and space requires a sensor distribution covering the whole sky. We discuss the potential of large-scale, small-platform-distributed architectures and constellations to build a sensitive X-ray/γ-ray all-sky monitor and the programmatic implications of this, including the set-up of an efficient assembly line for both hardware development and data analysis. We also discuss the potential of a constellation of small platforms operating at other wavelengths (UV/IR) that are capable of repointing quickly to follow-up high-energy transients.

**Keywords:** γ-ray burst; multi-messenger astrophysics; nano-satellites

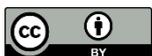

## 1. Introduction

The sky is teeming with explosive, energetic transient events, many of which remain hidden from our view. Some of the most exciting transient phenomena are γ-ray bursts (GRBs), discovered in 1967 by the Vela military satellites [1]. About two GRBs are detected per day, and last from a fraction of a second to a few minutes (in exceptional cases a few hours), in the energy range from several keV to a few MeV. They are some of the most extreme explosive events ever observed, momentarily outshining any other phenomena in the sky and are associated with the death of stars and the coalescence of compact objects (e.g., neutron stars) to form a new black hole. Despite great efforts and numerous observations, many open questions about their detailed physics remain. The emergence of multi-messenger astronomy provides a unique opportunity to shed new light on GRB physics [2]. To make progress, we need to perform a sensitive, all-sky monitoring of the high-energy sky, detect and localize the transients simultaneously with other probes, and follow them up rapidly with telescopes which observe other wavelengths.

It may sound like a mockery that the most dramatic events in the cosmos produce among the most luminous objects in the Universe (GRBs) but that all this light is most likely produced quite far from where the action is; far from the newborn event horizon, the accretion disk, and the region where relativistic jets are launched. On the other hand, gravitational waves (GWs), encoding the rapid/relativistic motion of compact objects, allow us to look directly into the innermost regions of these systems, providing precise information on space–time dynamics, and therefore the mass, spin, interior properties and inclination of the systems, as well as accurate distances. Electromagnetic measurements can hardly provide accuracies comparable to GW observations on these quantities, which are key to testing general relativity, the physics of compact objects and the build-up of the most efficient accelerators in the Universe. However, the information carried by GWs can be greatly amplified by identifying the context in which the event occurs. Electromagnetic observations can provide this context, as the GW/GRB170817 event strikingly demonstrated.

Multi-messenger astrophysics can include many more sources and astrophysical context, in addition to compact binary mergers, with their associated GRBs and kilonovae, such as supernovae, binary white dwarfs, coalescence of supermassive black holes, tidal disruption events in the vicinity of supermassive black holes and many others. In this paper, we limit ourselves to a discussion on compact binary coalescences (CBCs). The paper is organized as follows: we first discuss the main scientific goals of the multi-messenger approach to CBCs; we then summarize where we stand today; and, finally, discuss the role of distributed architectures in CBCs multi-messenger research during the present decade and the 2030s.

## 2. Why Multi-Messenger Astrophysics? What We Want to Learn from Compact Binary Coalescence

Binary systems including two compact objects (black holes, BH, and neutron stars, NS) are unique laboratories for studying physics that is inaccessible in terrestrial laboratories; in particular strong-field general relativity, relativistic acceleration mechanisms, and matter under extreme conditions (density, temperature, magnetic field). It was postulated for many years that the NS-NS and BH-NS coalescence leads to the production of a short GRB, a bright flash of gamma-rays lasting less than a few seconds [3] (see also the reviews) [4,5]. The reason for this can be outlines as follows. The outcome of the coalescence is a BH (or a metastable NS), including most of the mass of the binary system, and an accretion disk including the leftovers (on the order of a few percent of the solar mass). Accretion of this matter onto the newly formed compact object can release $10^{52}$–$10^{53}$ ergs of gravitational energy, so even if a small fraction of this large amount can be converted to electromagnetic radiation a GRB can be generated. The duration of the burst is determined by the lifetime of the disk, which is expected to be a fraction of a second at these masses. Accreting compact objects can produce powerful jets, which can transport the energy from the launching site near the inner engine to the photosphere, usually at $10^7$–$10^8$ Schwarzschild radii $R_s$. The jet launching site is thought to be near the newly formed BH, a potential laboratory for GR in the strong-field and the dynamics of matter under extreme conditions, or even new physics. In other words, are the BH and NS we observe in GW the same as described by GR? GWs are produced during the binary coalescence and settling of their remnants; therefore, they should precede the launching of the jet (the short GRB), while providing precise information on the physical condition of the system just before engine ignition. The physics at play outside the photosphere includes particle acceleration in internal shocks to produce the GRB prompt emission and external shocks to produce the GRB afterglow; both can be probed by electromagnetic observations. Detailed GRMHD simulations express this in a more quantitative framework ([6] and references therein). However, only observations can quantitatively confirm this scenario, connecting the physics near the event horizon of the newly formed BH to the physics of jet formation, collimation, and propagation. For example, the variability of the observed gamma-ray light curve may reflect the energy injection at the base of the jet [7,8]. The following key questions can be addressed by the multi-messenger approach to CBC:

- **What happens during the merger of compact objects?** How frequent is the coincidence with short GRBs; how frequent is the formation of powerful relativistic jets? Non-detections are, in principle, as important as detections. This question is addressed through simultaneous observations and studies of the GW event and the high-energy emission associated with jet production.
- **What is the nature of the short GRB's central engine**? What powers the most powerful accelerators of the Universe, NS or BH accretion? The study of the gravitational wave form can distinguish the nature of the remnant. The detection of a short GRB would indicate that a powerful accelerator is in place.
- **What is the jet launching mechanism?** The delay time between the GW emission and γ-rays emission (the short GRB) can distinguish between different jet launching scenarios [9].
- **Do jets have a universal structure or does the structure depend on the type of the compact binary?** Perhaps on the mass/spin of the binary components or of the merger remnant? The jet structure can be addressed through both the analysis of the prompt event and its afterglow [10], as well as direct VLBI imaging [11,12]. This requires accurate localizations and follow-up observations from radio to X-rays.
- **What is the role of CBCs in the production of heavy elements in the Universe?** This requires accurate localizations and follow-up spectroscopic observations from UV to IR.

The first three questions require the detection of GW signals from a statistical sample of NS-NS and NS-BH systems and their sensitive high-energy coverage over the full sky. The fourth and fifth questions require the accurate determination of the position of the source of GWs and multiwavelength follow-up observations, from radio waves to gamma-rays, of the electromagnetic counterpart.

## 3. CBC Multi-Messenger Astrophysics Today

It is often assumed that the multi-messenger revolution has a very precise starting date: 17 August 2017, with the detection of GW170817 by the LIGO/Virgo interferometers and short GRB170817 by the Fermi/GBM and INTEGRAL/SPI. Unfortunately, no other event was detected simultaneously by GWs and gamma-rays (or any electromagnetic radiation) during the third LIGO/VIRGO observing run (O3). Since one swallow does not make a summer, all questions presented in the previous section are still open. Certainly, GW/GRB1708017 has shown how powerful the multi-messenger approach could be. Its full impact, however, measured by the ability to open a long-lasting, brand-new field must still be determined, and it will greatly depend on the capability of collecting statistical samples of GW-electromagnetic events in the near future, that is, during the LIGO/VIRGO O4 and O5 observing runs planned for this decade.

As of now, the LIGO/Virgo interferometers have detected two NS-NS coalescence events (GW170817 and GW190425) and one very likely BH-NS merger (GW190814).

**The GW/GRB170817** event does not need to be commented on in detail here; many comprehensive reviews exist [6]. Briefly, Fermi/GBM and INTERGAL/SPI detected a short burst lasting for about 2 s, just 1.7 s after the NS-NS detection by LIGO/Virgo. The LIGO/Virgo error box was about 30 deg², a value that did not improve much after including the Inter Planetary Network (IPN) error box. The distance of the source $40^{+8}_{-14}$Mpc was determined directly from the GW detection and greatly limited the search for an optical counterpart to only about 50 galaxies. An optical counterpart was discovered in NGC4993 only after about 11 h from the GW/GRB event, which gave rise to an impressive follow-up by nearly one hundred ground-based and space-based telescopes, placing constraints on the optical/IR counterpart of the GW source/GRB and discovering the first confirmed kilonova in history. The GRB detection, which was nearly simultaneous with the GW signal, confirmed that, at least in this case, a merger of two NSs is the origin of short GRBs. The follow-up observations allowed the complex jet structure to be determined, and the jet was observed as off-axis for the first time [10–12]. It was also found that kilonovae were one of the prime sites of heavy element production through the r-processes [6,13].

**GW190425** was detected with a high significance by one interferometer only, and thus the uncertainty region included a large fraction of the sky [14]. No GRB was detected at the same time, with an upper limit on the 50–300 keV fluence of $\sim 10^{-6}$ ergs/cm² in 1 s [15], corresponding to a few ph/cm², by INTEGRAL/SPI. Swift/BAT and Fermi/GBM covered only a fraction of the region, and therefore could not contribute much to the INTEGRAL observation. The distance of the event inferred from the GW signal is $159^{+69}_{-71}$ Mpc. Therefore, even for a gamma-ray luminosity comparable to that of GRB170817, its flux would need to be 16 times dimmer, <0.1 ph/cm²; hence, it is relatively difficult to detect with the available detectors. Moreover, the observed gamma-ray flux is a strong function of the inclination at which the jet is observed (a power law with an exponent $-5 \div -6$). Unfortunately, the inclination of this binary system was not well constrained by the interferometers, but assuming similar characteristics of GRB170817, we can infer a lower limit on the inclination between the jet and the line-of-sight of about 10 deg.

**GW190814** was detected by all three interferometers, thus providing good constraints on the position of the source (error box of 18.5 deg²), distance ($241^{+41}_{-45}$ Mpc) and inclination (46+/−11 deg), in addition to the masses of the two coalescing objects and of the remnant (this was the GW event with the most unequal mass ratio, 0.11+\−0.01 ever detected). No significant electromagnetic counterpart was reported for this event, with INTEGRAL/SPI inferring a three-sigma upper limit for the flux of $3 \times 10^{-7}$ erg/cm²/s in the energy range 75–2000 keV [16]. Swift/BAT covered >99% of the error box with >10% partial coding, reporting a five-sigma upper limit of $10^{-7}$ erg/cm² between 15–350 keV in 1 s [17]. The lack of gamma-ray detection is likely due to the high inclination of the system, which corresponds to a huge (factor of $10^6$–$10^7$) reduction in the observed flux. The jet emission at these inclinations is hardly observable by any conceivable all-sky monitor at the moment.

In summary, the lessons learnt from the three NS-NS and BH-NS events collected so far are: (1) the need for an **all-sky monitor** with a sufficient sensitivity to detect off-axis jets with intrinsic luminosities down to $10^{47}$ ergs, at least to a distance of 100–200 Mpc and an inclination of 10–20 deg (of course smaller inclination values allow for detection over larger distances). In fact, within this decade the LIGO/Virgo interferometers will reach their target sensitivity and the searched volume will become much larger with respect to O1–O3: the horizon for NS-NS merging events detected with a signal-to-noise ratio of 8 will reach ~200 Mpc for LIGO and 100–130 Mpc for Virgo in O4, implying a discovery volume ~100 times larger than in the GW170817 case. The capability to instantaneously cover the whole sky is mandatory, because the number of events will be small; therefore, missing simply one event would mean a considerable loss for scientific research. (2) The capability of determining the position of the transients with uncertainties is smaller than a few degrees. Within the volume defined by this spatial constraint and the distance of the GWE provided by the interferometric measurements, the number of optical transients will be small enough to easily assess the correct transient to associate with the GWE, thus prompting further follow-ups.

## 4. The Role of Distributed Architectures in Tomorrow's Multi-Messenger Astrophysics

### 4.1. A Constellation of Nano-Satellites for High-Energy Transient Detection and Localization

Today, X-ray and gamma-ray monitors dedicated to the search and localization of high-energy transients are mostly monolithic instruments (e.g., NASA Swift/BAT, ESA INTEGRAL/IBIS, INTEGRAL/SPI, ASI AGILE/SuperAgile, AGILE/Microcalorimeter) or multiple detectors hosted by the same large spacecraft (NASA FERMI/GBM). One good example of distributed architecture used for GRB science since the beginning of this enterprise is the Inter-Planetary network (IPN). The accurate timing from X-ray and γ-ray detectors hosted on different spacecrafts for both Low-Earth Orbit (LEO) and High-Earth orbit (HEO) or even inter-planetary routes, is joined together to determine the transient position in the sky. This is achieved by using the delay time of arrival of the transient signal on different detectors, a strategy used since the earliest GRB detections by the VELA satellites at the end of the 1960s.

While the VELA satellites were fully dedicated to the detection of gamma-ray flashes (in this case the main target was the detection of gamma-rays from nuclear tests above the Earth's atmosphere), today, the IPN includes instruments on satellites fully dedicated to high-energy astrophysics, and instruments hosted by spacecrafts with quite different primary purpose. The diverse gamma-ray instrumentation used by the IPN, the poor knowledge of the spacecraft position when outside the Earth's GNSS infrastructure, as well as the difficulty of defining an absolute time with a good precision for all spacecrafts and the delay in communication with remote solar system spacecrafts, significantly limit the ability of the IPN to routinely provide accurate (better than a few degrees) and timely localizations. In other words, systematic errors associated with IPN transient localization are usually much greater than the statistical errors.

All of these difficulties could be overcome by a constellation of satellites in LEO hosting similar, if not identical, X-ray and gamma-ray detectors. The GPS and Galileo infrastructures offers the opportunity to constrain LEO satellite positions to within a few tens of meters, and the absolute time within a few tens of nanoseconds, orders of magnitude better than what is possible to achieve outside of the GNSS infrastructure. Adopting identical detectors ensures similar responses to cosmic events, reducing systematic uncertainties.

Since GRBs, and high-energy transients in general, are relatively bright (the flux from GRB170817 was about 2 ph/cm$^2$/s in the 50–300 keV band (and this was a faint GRB) seen from a relatively large off-axis jet angle), even a small instrument can be efficiently used for their detection. In fact, the collecting area of the Fermi/GBM modules is about 120 cm$^2$. Today, this class of instruments can be hosted by compact nano-satellites. Nano-satellites were developed at the end of the 20th century for didactical purposes, but today they are used for the most diverse applications: Earth observation, aircraft and ship tracking, telecommunication, and science. Nano-satellites have several major advantages with respect to traditional satellites. First and foremost, they can be developed on a relatively short timescale (a few years and, in extreme cases, a few months) compared with one to several decades usually required for standard space missions. Second, their cost is a few orders of magnitude smaller than standard large satellites. Both of these advantages imply that modularity can be exploited to its maximum. Modularity can allow us to (a) avoid single (or even multiple) point failures (if one or several units are lost the constellation and the experiment can still be operative); (b) fully test the hardware in orbit with the first launches and then improve it, if needed, with the following launches (iterative development, used in the context of gamma-ray flashes and GRBs early on by the VELA satellites. The second generation included gamma-ray detectors with much better timing capabilities, strongly improving the localization capabilities of the constellation); (c) to build the final mission step by step, gradually increasing its performance while diluting costs and risks.

An all-sky monitor, capable of monitoring the whole sky, or a large fraction of the sky, at all times, requires either a distributed infrastructure on LEO or a dedicated spacecraft far from the Earth. Given the advantages of miniaturized instrumentation hosted by CubeSats, it is relatively natural to propose building a sensitive all-sky monitor based on a constellation of nano-satellites.

In the previous section, in the context of multi-messenger astrophysics, we discussed the scientific relevance of even simple high-energy transient detections at the time of GWEs. Of course, their accurate localization would multiply the scientific return, prompting multi-wavelength (and even multi-messenger) follow-ups.

In the context of multi-wavelength astrophysics, it must be considered that, in a few years, the Vera Rubin Observatory (VRO) and the Cerenkov Array Telescope (CTA) will come online. VRO will have the ability to cover ~1/4 of the sky every night, finding millions of transients per night down to a magnitude r < 24.5 using real-time data analysis. On the one hand, optical counterparts of GRBs and GWEs will likely be serendipitously found in VRO images (covering each about 10 deg$^2$), providing arcsec positions and immediately prompting multiwavelength follow-up. On the other hand, the detection of X-ray and gamma-ray emission will promptly characterize the VRO transients (magnetars, soft gamma-ray repeaters, tidal distruption events, thermonuclear bursts from accreting NS, novae, AGN jets, etc., in addition to GRBs), thus better focusing the multiwavelength follow-up.

CTA will boost the study of the 20 GeV–300 TeV energy range. Only three GRBs have been detected at TeV energies so far: one by MAGIC [18] and two by HESS [19]. The CTA's fast re-positioning capabilities (20 s) and the improved sensitivity, due to the larger collecting area and lower energy threshold (~20 GeV) compared to MAGIC and HESS, will aid the study of GRB high-energy radiation routine, opening the possibility to accurately derive the jet Lorentz factor, assessing the role of synchrotron and inverse Compton radiation, and constraining the magnetic field strength and configuration. Given the limited field of view (FoV) of CTA at GeV energies (4.3°), an instrument operating during the 20s and providing the localization of GRBs with errors smaller than the CTA FoV is of paramount importance for triggering CTA follow-up observations.

*4.2. A Powerful Combination of Nano- and Micro-/Small Satellites*

Four years later, GW170817A remains the only gravitational wave source with a detection of an electromagnetic counterpart. While GW170817 was quickly followed by a short GRB seen at an angle of 19–42 degrees from the jet axis, it is likely that most kilonovae will not emit a GRB observable from the Earth. The prompt γ-ray emission is strongly beamed, and it is estimated that only about 1 in 100 kilonovae will be detectable at high energies [20]. However, GRB170817A is an unusually long and faint short-GRB, and detections of other GRB counterparts for gravitational wave events can be important discoveries. In the previous section, we discussed how a constellation of nano-satellites can be efficiently used to provide a powerful, high-energy, all-sky monitor at a relatively low cost and on short timescales. Here, we discuss how a synergic constellation of micro-satellites could be used for follow-up observations, greatly enhancing the scientific return.

It will require coordination, but to a large extent, several nano- and micro-satellite constellations with different detectors could work together in conjunction, forming one network. The micro-satellites will perform rapid follow-up observations at near-UV/optical and near-IR wavelengths. To follow up kilonovae without GRB counterparts, detected only by GW observatories, these observations shall be triggered directly by their GW emission. This will require the near UV/optical/IR observatories to have large fields of view and fast repointing capabilities, enabling them to locate the electromagnetic counterparts of kilonovae after short mosaicing observations. The early time evolution of the near-UV to near-IR flux ratios will provide the key diagnostics to distinguish between various scenarios of kilonova explosions. No existing or proposed mission provides all-sky monitoring and localization together with rapid multi-wavelength follow-up capabilities.

The rapid follow-up observations at near-UV/optical, near infrared, and X-ray wavelengths are expected to produce real breakthroughs in our understanding of kilonovae. The luminous optical counterpart of GW170817 was initially blue in colour with the emission peaking at near-UV wavelengths. Then, over the course of a few days the emission shifted to the near-IR wavelengths. This fast spectral evolution was unlike that of any previously observed event. However, the optical counterpart was discovered only about 11 h after the gravitational wave signal. A wide-field UV space telescope, able to rapidly slew on-source, could revolutionize our understanding of these exciting events (e.g.,

Ultrasat, http://space.gov.il/en/node/1129 10 December 2021). Theoretical modelling predicts that the first few hours might be dominated by near-UV emission from free neutrons, which do not have time to be captured by the nuclei. Observing this early emission is thus key for the understanding of the nucleosynthesis of kilonovae.

Figure 1 of Fernandez & Mezger 2016[21] presents the phases of a binary neutron star merger as a function of time, showing the observational signatures, as well as the possible outcomes and the associated physical phenomena. The in-spiral and coalescence of neutron stars, which can be observed through gravitational waves, can result in a hyper-massive neutron star that quickly collapses into a black hole; into a stable, rapidly spinning, highly magnetised neutron star; or directly into a black hole. The merger gives rise to the ejection of $10^{-4}$–$10^{-2}$ solar masses of unbound matter, with velocities 0.1–0.3 c from the tidal tails in the equatorial region [20]. The ejected matter that remains bound to the resulting compact object falls back and forms an accretion disc that helps launch the ultra-relativistic jet, which produces the observed short GRB. The equatorial ejecta are expected to be rich in heavier elements, known as lantanides, and produce long-lasting infrared emission. UV and blue emissions are produced early in the kilonova and last for only about a day. This may arise from free neutrons or from lantanide-poor polar ejecta with a higher electron fraction. While the properties of the tidal ejecta are sensitive to the mass ratio of the neutron stars, the properties of the polar and wind ejecta are sensitive to the neutron star radii and to the nature of the merger product. The different ratios between the observed kilonova fluxes obtained by near-UV, optical and near-infrared observations, will allow us to identify and constrain the properties of the different ejecta [13]. In particular, UV observations, performed in the first few hours of the kilonova (unavailable for GW170817), will probe the mass, composition, and thermal content of the fastest ejecta and allow us to constrain its geometry, quantity, and kinematics, as well as the nature of the merger product. Micro-satellites carrying relatively small (with a collecting area around 200 cm$^2$) near-IR and near-UV space telescopes will probe the emission from kilonovae out to the distance beyond 200 Mpc.

Observations in the near-IR should also allow us to detect afterglows at redshifts z > 5, corresponding to the first billion years of the Universe. Long GRBs are mostly observed at cosmological distances, with the most distant GRB detected at the redshift of z = 9.4, which corresponds to a look-back time of 13.3 billion years, only 500 million years after the Big Bang. Long GRBs are thus excellent probes for examining the early Universe, when the first massive stars and their host galaxies were being formed, the first heavier elements were produced, and the diffuse interstellar and intergalactic matter was re-ionised. To truly exploit long GRBs as probes of the early Universe, we need to identify more GRBs from the first billion years after the Big Bang. This can only be achieved by rapid near-infrared, follow-up observations from space, capable of imaging the afterglows of GRBs at the edge of the observable Universe. This was the main goal of the Theseus mission [22], which unfortunately was not selected by ESA for realization. CubSats again can provide a contribution here; see, for example, the SkyHopper proposal [23] https://skyhopper.research.unimelb.edu.au 10 December 2021.

Next to the near-IR and near-UV telescopes, the constellation would also benefit from a microsatellite carrying a wide-field 10 cm$^2$–20 cm$^2$ X-ray telescope observing in the 0.5–8 keV band. Since GRBs are bright in X-rays, a rapidly slewing X-ray telescope can aid the quick arcmin scale identification of the GRB position in the sky. The time and spectral evolution of the early X-ray emission can also provide valuable information about the possible two-step collapse model (through a short-lived massive neutron star) and the jet geometry of the source.

*4.3. An Incomplete List of Pathfinder/Precursor Missions*

In the past, many projects in the broad area of nano-satellites for high-energy astrophysics were proposed and several were funded, in order to build demonstrators or pathfinders. Among these, we mention the following, without pretending to be complete.

4.3.1. CAMELOT and GRBAlpha

One of the concepts developed to perform all-sky monitoring and timing-based localization of gamma-ray transients is called CAMELOT: Cubesats Applied for MEasuring and LOcalising Transients [24]. It is a constellation of 3U CubeSats equipped with large and thin (150 × 75 × 5 mm) CsI(Tl) scintillators read out by SiPM detectors, called multipixel photon counters (MPPCs), by Hamamatsu. The detectors are placed on two perpendicular walls of the satellites to maximize the effective photon collecting area on a CubeSat of this size.

The detector concept developed for the CAMELOT mission was first demonstrated on a 1U CubeSat, named GRBAlpha [25]. It carries a smaller, 75 × 75 × 5 mm CsI (Tl) scintillator, which provides one-eighth of the expected effective area of the 3U CubeSats envisioned for the CAMELOT mission. GRBAlpha was successfully launched on 22 March 2021 to a sun-synchronous polar orbit with a Soyuz 2.1 rocket from Bajkonur. Following a short commissioning phase, the detector was switched on and regular measurements began. The on-board data acquisition software stack is periodically updated in orbit, continuously improving the capabilities of the science payload. The degradation of the SiPM detectors, which are protected by a 2 mm thick lead shield, is being monitored. About half of the polar orbit is plagued by a high particle background, and thus the duty cycle of the detector cannot be better than 50%, even if it is operated continuously. The satellite communicates on amateur radio frequencies and its ground segment is supported by the radio amateur community. The mission also takes advantage of the SatNOGS network for increased data downlink volume. Figure 1 shows GRBAlpha before the launch.

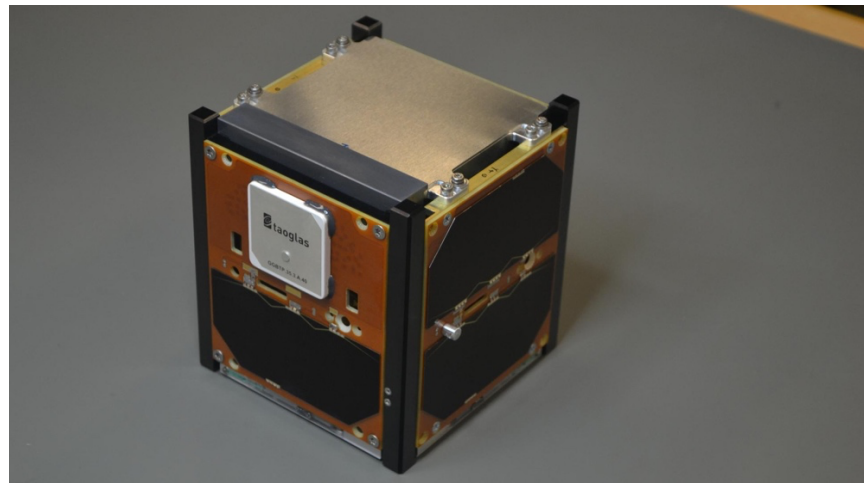

**Figure 1.** GRBAlpha, a 1U CubeSAT equipped with a gamma-ray detector (visible on the top, see [25] for details).

GRBAlpha became the first nano-satellite to detect multiple confirmed GRBs. At the time of writing, the mission detected five GRBs (four long and one short GRBs [26–30]), two of them over the course of a single night from 18 to 19 October. Figure 2 shows the light curves of GRB211018A observed by GRBAlpha in different energy bands [30]. GRB detectors developed for the CAMELOT mission are also going to be launched on VZLUSAT-2, which is a technological 3U CubeSat built by the Czech Aerospace Research Centre. The satellite will carry two perpendicular detectors the same size as GRBAlpha.

VZLUSAT-2 is expected to be launched in January 2022 into a sun-synchronous polar orbit on a Falcon 9 rideshare mission.

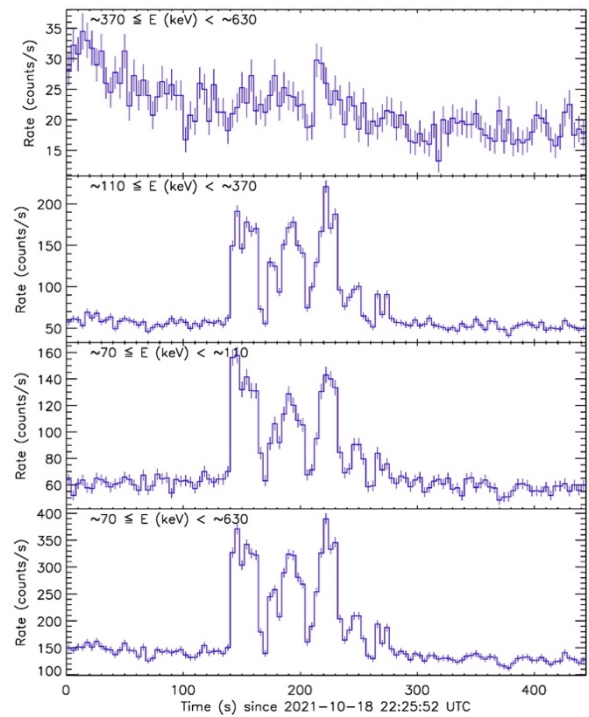

**Figure 2.** Light curves of GRB 211018A observed by GRBAlpha in different energy bands [30].

4.3.2. HERMES Pathfinder

HERMES-Technologic and Scientific pathfinder (HERMES pathfinder) is an in-orbit demonstration consisting of a constellation of six 3U nano-satellites hosting simple but innovative X-ray detectors for the monitoring of cosmic high-energy transients such as GRBs and the electromagnetic counterparts of GWEs [31]. The main objective of HERMES-TP/SP is to prove that an accurate position of high-energy cosmic transients can be obtained using miniaturized hardware, with a cost at least one order of magnitude smaller than that of conventional scientific space observatories and a development time as short as a few years.

The transient position is obtained by studying the delay time of arrival of the signal to different detectors hosted by nano-satellites on low Earth orbits [32]. To this purpose, particular attention is placed on reaching the best time resolution and time accuracy, with the goal of reaching an overall accuracy of a fraction of a micro-second [33]. The main goals of the project are: (1) join the multi-messenger revolution by providing the first mini-constellation for GRB localization with a total of six units (the first experiment of GRB triangulation with miniaturized instrumentation); (2) develop miniaturized payload technology for breakthrough science; (3) demonstrate COTS applicability to challenging missions, contribute to Space 4.0 goals, push and prepare for high-reliability large constellations.

Figure 3 shows the HERMES pathfinder detector system during integration. The 60 GAGG scintillator crystals can be seen to the right and the 12 $10 \times 10$ silicon drift detector mosaics used to read out the crystals to the left (see [33] and references therein for a detailed description of the HERMES pathfinder payload).

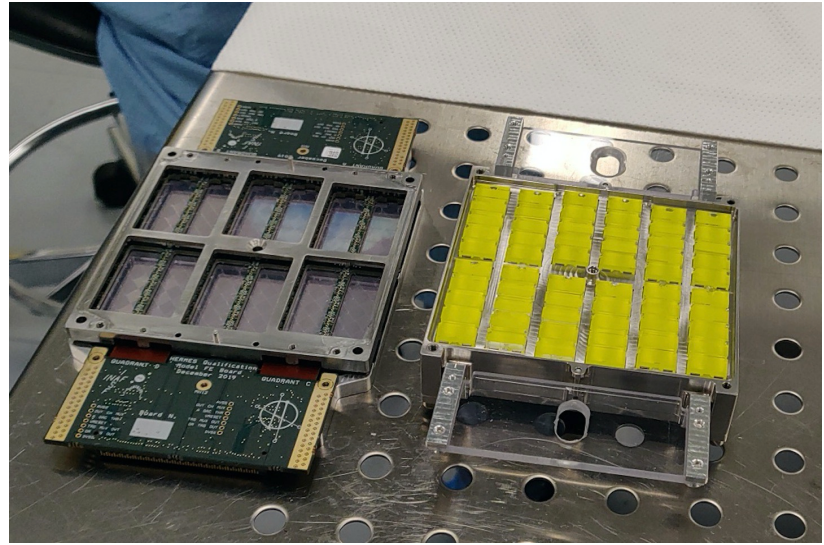

**Figure 3.** HERMES pathfinder detector system during integration at Fondazione Bruno Kessler laboratories in Trento, Italy.

The HERMES-TP project is funded by the Italian Ministry for Education, University and Research, and the Italian Space Agency. The HERMES-SP project is funded by the European Union's Horizon 2020 Research and Innovation Programme under Grant Agreement No. 821896.

The consortium started the integration and testing of the first flight unit during the summer of 2021; the proto-flight model and its qualification review is foreseen for Q1 2022. The other five units will be integrated and tested during 2022 and the constellation is set to be launched to a nearly equatorial LEO in 2023.

HERMES pathfinder is intrinsically a modular experiment that can be naturally expanded to provide a global, sensitive, all-sky monitor for high-energy transients. The next step is SpIRIT, a 6U cubesat funded by the Australian Space Agency and managed by the University of Melbourne. SpIRIT will host one HERMES pathfinder payload, and will fly on an SSO at the same time as the HERMES pathfinder, forming a constellation of seven satellites in two different orbits.

### 4.3.3. GALI

A new concept for identifying the direction of GRBs was suggested recently by Rahin et al. [34]. The concept was named GALI (GAmma-ray burst Localizing Instrument). Its basic idea is to use numerous small scintillators (e.g., $1 \times 1 \times 1$ cm$^3$ cubes) in a 3D array utilizing their mutual shielding. Consequently, the relative γ-photon count of each scintillator varies strongly with the direction of the burst. In a sense, GALI can be thought of as a coded-mask detector, but where the mask itself has detecting elements. Moreover, the detector (and mask) have no preferred direction, and thus provide full-sky coverage, as opposed to coded-mask instruments. A configuration such as CAMELOT, benefits from the SiPMs, which occupy little volume; hence, they enable the compact packing of the scintillators. As with GRBAlpha/CAMELOT, the SiPMs are radiation-sensitive, and need to be protected. A GALI laboratory prototype was successfully tested, and a flight model to be launched to the International Space Station (ISS) is being built. The laboratory prototype is shown in Figure 4.

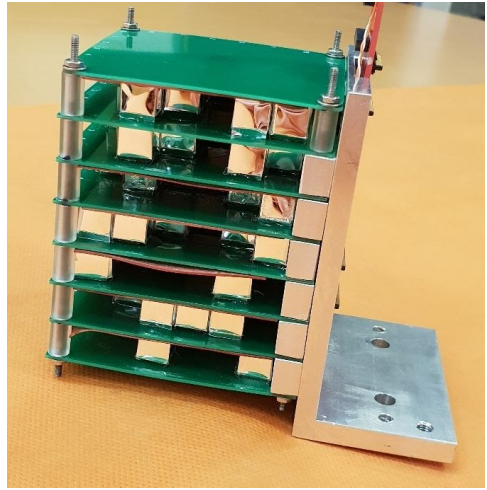

**Figure 4.** Laboratory model of the GALI detector system with 90 scintillators crystals. The individual scintillators in their reflective wrappings can be seen, arranged in a random order. A larger version is currently being built for the ISS.

The GALI concept can be scaled to any size, and thus can fit many platforms. Clearly, larger versions will be more sensitive, and, more importantly, they will provide a better angular resolution. An advantage of the GALI configuration is the reduced sky background on the inner scintllators, which will light up only for GRBs in specific directions. This allows for a high signal-to-background ratio on these scintillators, and the exploitation of soft γ-rays below 50 keV for directionality. The aforementioned flight model will consist of 350 scintillators, occupying a total volume of merely ~1 L. Simulations show that even such a small detector can identify the direction of a burst down to approximately ±2° for 1 s GRBs with a 10 keV–1 MeV flux of 10 ph cm$^{-2}$ s$^{-1}$, ±5° for 5 ph cm$^{-2}$ s$^{-1}$, and ±10° for 2.5 ph cm$^{-2}$ s$^{-1}$ [34]. Although GALI can operate onboard a single satellite, it can also be incorporated into a distributed satellite architecture to enhance the sky coverage and directional capabilities of the entire constellation.

4.3.4. Other Projects

BurstCube is a 6U CubeSat developed by NASA, which will detect GRBs using four CsI scintillators, each with an effective area ~90 cm$^2$ [35]. BurstCube is expected to be launched in 2022.

The Educational Irish Research Satellite 1 (EIRSAT-1), supported by ESA's Fly Your Satellite program, will carry a gamma-ray module (GMOD) to detect gamma-ray bursts [36]. GMOD uses SensL B-series SiPM detectors and a CeBr scintillator. EIRSAT-1 will be launched from the ISS in 2022.

Nanosatellite constellations include the Chinese Gamma-Ray Integrated Detectors (GRID), which will consist of GRB detectors (as secondary payloads) on 10–24 CubeSats [37]. Two GRID units have been launched so far and one GRB has been recently detected by the second unit [38].

**Author Contributions: conceptualization, F.F., N.W., E.B, writing, F.F., N.W., E.B.** All authors have read and agreed to the published version of the manuscript.

**Funding:** This research was funded by European Union Horizon 2020 Research and Innovation Framework Programme under grant agreements HERMES-SP n. 821896 and AHEAD2020 n. 871158, by ASI INAF Accordo Attuativo n. 2018-10-HH.1.2020 HERMES—Technologic Pathfinder Attivita' scientifiche, by a Center of Excellence of the Israel Science Foundation, grant No. 2752/19, and by a grant from the Israel Space Agency.

**Institutional Review Board Statement:** Not applicable.

**Informed Consent Statement:** Not applicable.

**Data Availability Statement:** Not applicable

**Acknowledgments:** This article benefited from the invaluable input of the HERMES pathfinder, CAMELOT and GALI teams. In particular we would like to thank J. Ripa, A. Sanna, L. Burderi, M. Lavagna, P. Bellutti, Y. Evangelista, M. Trenti for inspiring discussions.

**Conflicts of Interest:** The authors declare no conflict of interest.